\documentclass[11pt]{article}
\usepackage{amssymb}
\usepackage{amssymb,amsmath}
\usepackage{graphicx}
\usepackage{flafter}
\def\F{{\cal F}}

\def\diag{{\rm diag\,}}

\def\p{\partial}

\newcommand{\be}{\begin{equation}}
\newcommand{\ee}{\end{equation}}
\newcommand{\bd}{\begin{displaymath}}
\newcommand{\ed}{\end{displaymath}}
\newcommand{\ba}{\begin{array}{ll}}
\newcommand{\ea}{\end{array} }
\newcommand{\baa}{\begin{eqnarray}}
\newcommand{\eaa}{\end{eqnarray}}
\newcommand{\baaa}{\begin{eqnarray*}}
\newcommand{\eaaa}{\end{eqnarray*}}
\date{ }
\title{
Explicit formulas for currents at branching long lines and for
maximum of current amplitudes \footnote{IEE Proceedings - A, Vol.
140, No. 4, July 1993, pp. 249-251.}}
\author{
N.G. Dokuchaev, PhD}
\begin{document}\vspace{-0.5cm}      \maketitle
\begin{abstract}
 A system of 'telegrapher's' equations for a number of long lines
 joined into a network is studied. Explicit formulas for Fourier
 transforms of current and voltage are derived. These formulas are
 very suitable for computer application as well as for the
 analytical study of processes o networks. As an example, the
 availability of formulas aids the derivation of explicit formulas
 for maxima of current amplitude over the given class of admissible
 external influences. These values may be used to indicate the
 characteristic of network robustness to excess voltage or
 electromagnetic impulse. The approach is based on the operational
 solution already proposed by the author for more general partial
 differential equations on graphs.
\\
  {\bf Indexing terms}:
  Telegrapher's equations, Branching long lines, Waves on graphs
\end{abstract}
\section{Introduction}
The classical 'telegrapher's' equation describes the evolution of
current in a long line  [1]. There are a lot of technical and
physical objects that are modeled by a system (network) of jointed
long lines( a network of underground pipes, a system of grounding,
an ordinary electrical network of two-wire transmission lines with
consumers and sources etc.). In this case the 'telegraph equations'
are joined into a system with corresponding boundary value
conditions. All the existing methods of calculating the currents in
these systems are based on finite-dimensional approximation of the
continuous long lines(finite-elements methods, etc.), and so there
is a loss of precision in this approximation. These methods usually
need a prior manual analysis of the network topology. For some
particular kinds of problem(when interaction between the processes
at the different branches is realized only in a form of Kirchhoff's
law at the nodes where branches are connected), we obtain a new
method based on the approach [2] for the partial differential
equations on graphs. We derive explicit formulas for the Fourier
transforms of the currents and voltages in the branches of the
branching network. These formulas give an exact solution of the
system of 'telegrapher's equations'; the corresponding algorithm is
suitable for computer applications. The input data are provided
directly by the network topology, without the necessity for prior
manual analysis to obtain additional equations for boundary-value
description. The availability of the formulas for the currents is
used to derive the explicit formulas for currents amplitude maximum
being achieved over all the external influences (inputs) satisfying
the constraints for input energy or input value at every point. This
maximum may be used as a characteristic of network robustness to
excess voltage.
\par
Let us consider a system of $n$ long lines with lengths
$l_1,\dots,l_n$ that are connected to a network (graph with branches
and nodes). The equation for the current and voltage in each branch
with number $k=1,\dots, n$ is [1] \begin{align}
\begin{cases} \frac{\p U_k}{\p y}(y,\omega) &= -(R_k + i \omega
I_k)I_k(y,\omega) +
e_k(y,\omega)\\
\frac{\p I_k}{\p y}(y,\omega) &= -(Y_k + i \omega C_k)U_k(y,\omega)
+ \delta (y-\bar{y}_k)\bar{I}_k(\omega)
\end{cases}
\end{align}
In eqn.1, $I_k(y,\omega)$, $U_k(y,\omega)$ are the Fourier
transforms of current and voltage in the $k$th branch, $y\in[0,l_k]$
is a longitudinal co-ordinate (we have chosen the orientation for
every branch), $\omega$ is an angular frequency, $\omega\in
\mathbb{R}$. The constants $R_k$, $I_k$ are the series resistance
and inductance per unit length. The constants $Y_k$, $C_k$ are the
parallel conductance and capacitance per unit length. The action of
the external electric field is characterized by the functions
$e_k(y,\omega)$; $\bar I_k(\omega)$ are the external currents
flowing into the given points $\bar{y}_k\in [0, I_k]$; and $\delta $
is a delta function.
\par
    We assume that interaction between the processes at the
    different branches is realized only at the nodes where branches
    a re connected, and so all the eqns. 1 for $k=1,\dots, n$ are
    joined into the system with supplementary boundary-value
    conditions in the form of Kirchhoff's current and voltage laws
    at joining nodes. Ohm's law holds at the terminal nodes that
    belong to only one branch, so that $I_k(0,\omega) =\bar{Y}_kU(0,\omega)$
    or $I_k(I_k, \omega)=\bar{Y}_kU_k(I_k,\omega)$, where the
    constants $\bar{Y}_k$ are the terminating admittances.
\par
    We suppose that
\par
  \renewcommand{\labelenumi}{(\roman{enumi})}
\begin{enumerate}
\item
Nonzero $R_k$ or $Y_k$ exist for some $k$.
\item
    $|e_k(y,\omega)|< c$ for some constant $c>0$ for every $k$, $y$,
    $\omega$.
\end{enumerate}
\section{Formulas for currents and voltages}
Our method is taken from Reference 2. The main idea is to describe
the current and voltage distribution by the complex $2n$-vector
\begin{align}
 z(x,\omega)= [U_1(l_1 x, \omega),I_1(l_1 x, \omega),
 \dots, U_n(l_n x, \omega),I_n(l_n x, \omega)]^\top
\end{align}
defined for every $x\in [0,1]$,$\omega \in \mathbb{R}$. This vector
is convenient because we can write the boundary value problem for
it: \baa
    &&\frac{\p z}{\p x}(x, \omega) = A(x,\omega)z(x,\omega)+
    E(x,\omega),\nonumber\\
    &&B_0z(0,\omega) + B_1 z(1,\omega)=0.
 \eaa
 Here the external influence is
\begin{align}
    & E(x,\omega) = [l_1e_1(l_1,\omega),\delta(l_1x-\bar{y}_1)\bar{I}_1(\omega),
    \dots, e_n(l_n x, \omega), \delta (l_n x -
    \bar{y}_n)\bar{I}_n(\omega)]^\top
\end{align}
In eqn.3, $A(\omega) = \diag[A_1(\omega), \dots, A_n(\omega)]$ is a
complex $2n\times 2n$-matrix, where $2\times2$ matrices
\begin{align*}
     A_k(\omega) = l_k
     \begin{pmatrix}
     0 & -R_k -i\omega L_k\\
     -Y_k -i\omega C_k & 0
     \end{pmatrix}
     ,  k = 1,\dots, n
\end{align*}
This means that the matrices $A_k$ are placed on the main diagonal
of $A$ and all the remaining elements of $A$ are zero.
\par

The $2n\times2n$-matrices $B_0$ and $B_1$ contain complete
information about the topology of our network; we shall give the
algorithm for their construction below:
\par
The explicit final formula for the Fourier transform of the current
and voltage is
\begin{align}
    z(x,\omega) = \int_0^1 G(x,\rho, \omega)E(\rho, \omega)d\rho
\end{align}
together with eqn.2.

\par
    In eqn.5, $G(x,\rho, \omega)$ is the Green's function for the
    problem of eqn.3 defined [3] for $x\in[0,1]$, $\rho\in[0,1]$, $\omega \in \mathbb{R}$
    as a complex $2n\times 2n$ matrix
\begin{align}
    G(x,\rho, \omega) =
    \begin{cases}
        -\Phi(x,\omega)[B_0 + B_1\Phi(1,\omega)]^{-1}B_1\Phi(1-\rho,
        \omega) \text{  for } 0\leq x < \rho\\
 \Phi(x,\omega)[B_0 + B_1\Phi(1,\omega)]^{-1}B_0\Phi(-\rho,
        \omega) \text{  for } \rho< x \leq 1
    \end{cases}
\end{align}

Note that the $2k$th component of eqn.5 is undefined at the point
$x=l_{2k}^{-1}\bar{y}_{2k}$, and has a jump discontinuity at this
point for $k=1,\dots, n$ (see eqn.1).

\par
    The matrix $\Phi(x,\omega)=\exp[A(\omega)x]$ is the fundamental
    matrix of the first eqn.3: $\Phi(x,\omega)=\diag[\Phi_1(x,\omega),
    \dots,\Phi_n(x,\omega)]$, where
\[
    \Phi_k(x,\omega) =
    \begin{pmatrix}
        \cosh (\gamma_k x) & -\frac{\gamma_k}{\zeta_k}\sinh (\gamma_k
        x)\\
         -\frac{\gamma_k}{\xi_k}\sinh (\gamma_k
        x) & \cosh (\gamma_k x)
    \end{pmatrix}
\]
$\xi_k=l_k(R_k+i\omega L_k)$, $\zeta_k = l_k(Y_k+i\omega C_k)$,
$\gamma_k = \sqrt{\xi_k \zeta_k}$ (we take the principal value for
the root and assume that $\sinh 0/0 =1 $ ).
\par
    The topology of the network is defined by the ordered set $\{s_m, p_m,r_m,
    (i_1,\dots,i_{p_m}),(j_1,\dots,j_{r_m})\}_{m=1}^{q}$. Here $q$
    is the number of nodes, $p_m$ is the number of branches with
    numbers $i_1,\dots,i_{p_m}$ existing from the $m$th nodes, $r_m$
    is the number of the branches with numbers $j_1,\dots, j_{p_m}$
    entering into the $m$th nodes, $s_m = p_m + r_m$.
\par
    We construct the matrices $B_0$ and $B_1$ by joining all the
    strings of $s_m \times n$-matrices $Q_m = Q_m(i,j)$, $P_m =
    P(i,j)$,$m=1,\dots,q$, $i=1,\dots,s_m$, $j=1,\dots,n$:
\[
    B_0 = \begin{vmatrix}
    P_1\\
    \vdots\\
    P_q
    \end{vmatrix},\qquad B_1= \begin{vmatrix}
        Q_1\\
        \vdots\\
        Q_q
    \end{vmatrix}
\]
    where $Q_m$ and $P_m$ are constructed by the following
    algorithm.
\begin{enumerate}
\item
    Let $s_m=1$. If $p_m=1$, then $P_m(1,2i_1-1)=\bar{Y}_k$ and
    $P_m(1,2i_1)=-1$; if $r_m=1$ then $Q_m(1,2j_1-1)=\bar{Y}_k$ and
    $Q_m(1,2j_1)= -1$. All the remaining components of $P_m$ and $Q_m$
    must be zero.
\item
   Let $s_m>1$. If $r_m>0$, then $Q_m(p_m + d, 2j_d-1)=-Q_m(p_m+d,2j_{d+1}-1)=1$
   for $d=1,\dots,r_m-1$, and $Q_m(s_m,
   2j_1-1)=Q_m(s_m,2j_2)=\dots=Q_m(s_m,2j_{r_m})=1$. If $p_m>0$,
   then $P(s,2i_s-1)=-P_m(s,2i_{s+1}-1)=1$ for $s=1,\dots,p_m-1$ and
   $P_m(s_m,2i_2)=\dots=P_m(s_m,2i_{p\infty})=1$. If $r_m>1$ and $
   p_m>1$, then $P_m(p_m,2i_{pm}-1)=-Q_m(p_m, 2j_1-1)=1$. All the
   remaining components of $P_m$ an $Q_m$ must be zero.
\end{enumerate}
    Let us discuss the correctness of our formulas. The existence of
    nonzero $R_k$ or $Y_k$ is sufficient for matrix $[B_0+B_1\Phi(1,\omega)]$
    to be invertible, and for the problem of eqn.3 to be well
    posed[3]. Let us show this. Let $z_0\in\mathbb{C}^n$ and
    $[B_0 + B_1\Phi(1,\omega)]z_0 = 0$. There is an energy
    dissipation in eqn.1 for $k$ with nonzero $R_k$ or $Y_k$, and so
    the corresponding evolution system is stable in the temporal
    domain. Thus $z_0=0$, and the problem in eqn.3 is well posed.
\par
    Integral eqn.5 exists because of supposition (ii).
\section{Maximum of current amplitude}

Having obtained explicit formulas for currents, we can derive the
explicit formulas for the maximum current amplitude at a branching
network. This problem is important for applications, because we can
use this value for the characterization of the robustness to excess
voltage or electromagnetic impulse. We think that this problem has
no explicit solution other than our explicit formulas for currents.

\par

In eqn.1, the external influences are characteristised for every
$\omega \in \mathbb{R}$ by the ordered set
$F(\omega)=\{e_k(.,\omega), \bar{I}_k(\omega),\bar{y}_k\}_{k=1}^n$.
\def\F{\cal F}
\par
Fix $a_k\geq 0$,$b_k\geq0$, $0\leq\alpha_k\leq\beta_k\leq l_k$ for
$k=1,\dots,n$. Let us introduce the sets ${ F}_p(\omega)=\{{\cal
F}(\omega)\}$ being the classes of admissible external influences:
\begin{enumerate}
\item
 For $1\leq p< + \infty$, class ${ F}_p(\omega)$  is the set of such
$\F(\omega)=\{e_k(.,\omega), \bar{I}_k(\omega),\bar{y}_k\}_{k=1}^n$
that
\[
    \left(\int_0^{l_k}|e_k(y,\omega)|^pdy\right)^{1/p}\leq
    a_k,\quad
     |\bar{I}_k(\omega)|\leq b_k,\quad \bar{y}_k\in[\alpha_k,\beta_k],\quad k=1,\dots,n.
\]
\item
 Class $F_{\infty}(\omega)$ is the set of such $\F(\omega)=\{e_k(.,\omega),
 \bar{I}_k(\omega),\bar{y}_k\}_{k=1}^n$ that
 \[
    \sup_{y\in[0,l_k]}|e_k(y,\omega)|\leq a_k,\quad
    |\bar{I}_k(\omega)|\le b_k,\quad \bar{y}_k \in
    [\alpha_k,\beta_k],
    \quad k=1,\dots, n.
 \]
\end{enumerate}
Our aim is to obtain the value
\[
\sup _{\F(\omega)\in F_p(\omega)}|I_k(y, \omega)|
\]
for every $k=1,\dots, n$, $y\in[0,l_k]$, $1\leq p \leq +\infty$.
\par
The result is that, for $p\in (1,+\infty]$, we have
\begin{multline}\label{7}
\sup_{\F(\omega)\in F_p(\omega)}|I_k(y,\omega)| = \sum_{m=1}^n
\\ \left\{ a_k\left[\int_0^1|G_{2k,2m-1}(y/l_k,\rho,\omega)|^q
d\rho\right]^{1/q} + b_k \sup_{\rho\in[\alpha_k/l_k,
\beta_k/l_k]}|G_{2k,2m}(y/l_k,\rho,\omega)| \right\}
\end{multline}
where $q=p(p-1)^{-1}$ for every $p\in(1,\infty)$, $q=1$ for
$p=+\infty$. For $p=1$, we have
\begin{multline}
\sup_{\F(\omega)\in F_1(\omega)}|I_k(y,\omega)| = \sum_{m=1}^n
\\ \left\{ a_k \sup_{\rho\in[0,1]}|G_{2k,2m-1}(y/l_k,\rho,\omega)|+ b_k
\sup_{\rho\in[\alpha_k/l_k,
\beta_k/l_k]}|G_{2k,2m}(y/l_k,\rho,\omega)|\right\}
\end{multline}
Here, $G_{k,m}$ are components of matrix $G$ defined by eqn.6.
\par
Eqns.7 and 8 are very suitable for computer application. Analogous
results can be obtained for
\[
    \sup_{\F(\omega)\in F_p(\omega)}|U_k(y,\omega)|
\]
(the corresponding changes in eqns.7 and 8 are obvious).
\par
    To prove eqns.7 and 8, we have to remark that the right-hand
    part of eqn.7 or eqn.8 is the functional norm of the $2k$th
    string of $G(y/l_k,\cdot,\omega)$ being presented as an element
    of a Banach space [4], which is dual to such space of function $e_k(\cdot,\omega)$
    that $\F(\omega)\epsilon F_p(\omega)$ defines a ball there.
\section*{References}
$\hphantom{xk}$ [1] WEBER, E.: 'Linear transient analysis. Vol.
2'(Wiley and Sons, New York, 1956)
\par
[2] DOKUCHAEV, N.G.: 'Operation method for the boundary value
problems on graphs', {\em Differential Equations
[Differentzial'nyeuravnenia],} 1990, 26,(11), pp.2006-2008(in
Russian)
\par
[3] KRENER, A.J.: 'A causal realization theory. Part 1: Linear
deterministic systems', {\em SIAM J. Control}, 1987,25,(3),
pp.499-525
\par
[4] RUDIN, W.: 'Functional analysis' (McGraw-Hill, New York, 1973)
\end{document}